\documentclass[conference]{IEEEtran}
\usepackage[absolute,showboxes,overlay]{textpos}
\usepackage{hyperref}

\TPMargin{5pt}

\newcommand{\copyrightstatement}{
    \begin{textblock}{0.84}(0.08,0.93)    
         \noindent
         \footnotesize
         \copyright 2021 IEEE. Personal use of this material is permitted. Permission from IEEE must be obtained for all other uses, in any current or future media, including reprinting/republishing this material for advertising or promotional purposes, creating new collective works, for resale or redistribution to servers or lists, or reuse of any copyrighted component of this work in other works. {Cite from IEEE: }\href{<https://ieeexplore.ieee.org/document/9653043>}{DOI No. 10.1109/MILCOM52596.2021.9653043}
    \end{textblock}
}

\IEEEoverridecommandlockouts
\usepackage{cite}
\usepackage{amsmath,amssymb,amsfonts}
\usepackage{graphicx}
\usepackage{textcomp}
\usepackage{xcolor}

\usepackage{subfigure}
\usepackage{fancyhdr}
\usepackage{algorithm}
\usepackage{algpseudocode}
\usepackage{booktabs} 
\usepackage{multirow}

\include{pythonlisting}

\def\BibTeX{{\rm B\kern-.05em{\sc i\kern-.025em b}\kern-.08em
    T\kern-.1667em\lower.7ex\hbox{E}\kern-.125emX}}
\begin{document}
\copyrightstatement  

\title{A Semi-Supervised Learning Approach for Ranging Error Mitigation Based on UWB Waveform}

\author{
\IEEEauthorblockN{
Yuxiao~Li\IEEEauthorrefmark{1},
Santiago~Mazuelas\IEEEauthorrefmark{2}, and
Yuan~Shen\IEEEauthorrefmark{1}}
\IEEEauthorblockA{\IEEEauthorrefmark{1}
Department of Electronic Engineering,
Tsinghua University,
Beijing, China \\
}
\IEEEauthorblockA{\IEEEauthorrefmark{2}
BCAM-Basque Center for Applied Mathematics, and IKERBASQUE-Basque Foundation for Science, Bilbao, Spain \\
Email: li-yx18@mails.tsinghua.edu.cn,
smazuelas@bcamath.org,
shenyuan\_ee@tsinghua.edu.cn
}}

\maketitle

\begin{abstract}

Localization systems based on ultra-wide band (UWB) measurements can have unsatisfactory performance in harsh environments due to the presence of non-line-of-sight (NLOS) errors. Learning-based methods for error mitigation have shown great performance improvement via directly exploiting the wideband waveform instead of handcrafted features. However, these methods require data samples fully labeled with actual measurement errors for training, which leads to time-consuming data collection. In this paper, we propose a semi-supervised learning method based on variational Bayes for UWB ranging error mitigation. Combining deep learning techniques and statistic tools, our method can efficiently accumulate knowledge from both labeled and unlabeled data samples. Extensive experiments illustrate the effectiveness of the proposed method under different supervision rates, and the superiority compared to other fully supervised methods even at a low supervision rate.

\end{abstract}

\begin{IEEEkeywords}
Variational Bayes, Deep Learning, Semi-Supervised Learning, UWB Radio, ranging error mitigation
\end{IEEEkeywords}

\section{Introduction}
\label{sec:intro}

Wireless network localization is a key enabler for a wide range of emerging applications with the requirement of high-accuracy positional information \cite{WinSheDai:J18,WinDaiShe:J18}. 
Many papers develop localization algorithms based on range-related measurements. 
Among the measuring approaches for these algorithms, ultra-wideband (UWB) radio emerges to be a dominant trend due to the fine delay resolution and obstacle-penetration \cite{WymMarGifWin:J12}. However, its practical performance for range measurements is greatly degraded in harsh environments due to 
multipath effects \cite{WinChrMol:J06}, and non-line-of-sight (NLOS) conditions \cite{JohShuPet:J07}.

Multiple ranging error mitigation techniques are proposed to improve the distance estimates based on measurements, and subsequently the localization accuracy. 
Conventional methods are mostly model-based, with a simplified model for signal propagation mechanism \cite{KhaKarMou:J10}. 
Early learning-based methods, such as Support Vector machine (SVM) \cite{MarGifWymWin:J10}, 
formulate mitigation as a regression problem from statistical characteristics of measurements to ranging errors. 
These characteristics, however, require a time-consuming extraction phase and may still lose information inherent in the raw waveform \cite{MazConAllWin:J18,ConMazBar:J19}.

Deep learning methods represent to be a popular trend and have also been applied to wireless communications \cite{ZhaYueKat:C17}. Benefiting from a more thorough exploitation of raw data, these methods show great improvements in effectiveness and efficiency compared to conventional methods \cite{AngMazSalFanChi:J20,LiMazShe:C22,LiMazShe:C22_2}. 
However, these methods require a large amount of fully labeled data to achieve satisfactory results, leading to not-efficient work for data collection.



In this paper, we propose a semi-supervised method based on variational Bayes (VB) for the UWB ranging error mitigation problem, referred to as Semi-VL. Specifically, the generation process of received waveform is assumed to consist of two latent variables: one for range-related features and the other for unrelated environment semantics. 
The loss function, derived from variational inference, is composed of a supervised term and an unsupervised term. As a result, the proposed method can efficiently exploit information from both labeled and unlabeled data. Extensive experiments illustrate that the proposed Semi-VL achieves efficient error estimation under a range of supervision rates, and provides significant performance improvement compared to conventional fully supervised methods even at a low supervised rate.


\section{Variational Bayesian Method}
\label{sec:method}

In this section, we propose a variational Bayesian method to extract specific features from raw received signals for the ranging error mitigation problem. 

\subsection{Generative Model}


\begin{figure}[bp]
    \begin{center}
        \subfigure[]{
        \begin{minipage}[t]{0.25\linewidth}
        \centerline{\includegraphics[width=1.0\textwidth]{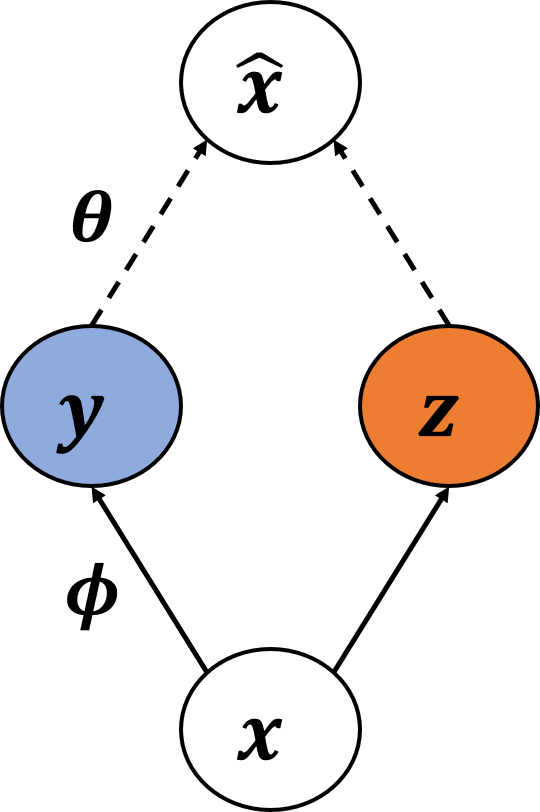}}
        \end{minipage}%
        }
        \subfigure[]{
        \begin{minipage}[t]{0.65\linewidth}
        \centerline{\includegraphics[width=1.0\textwidth]{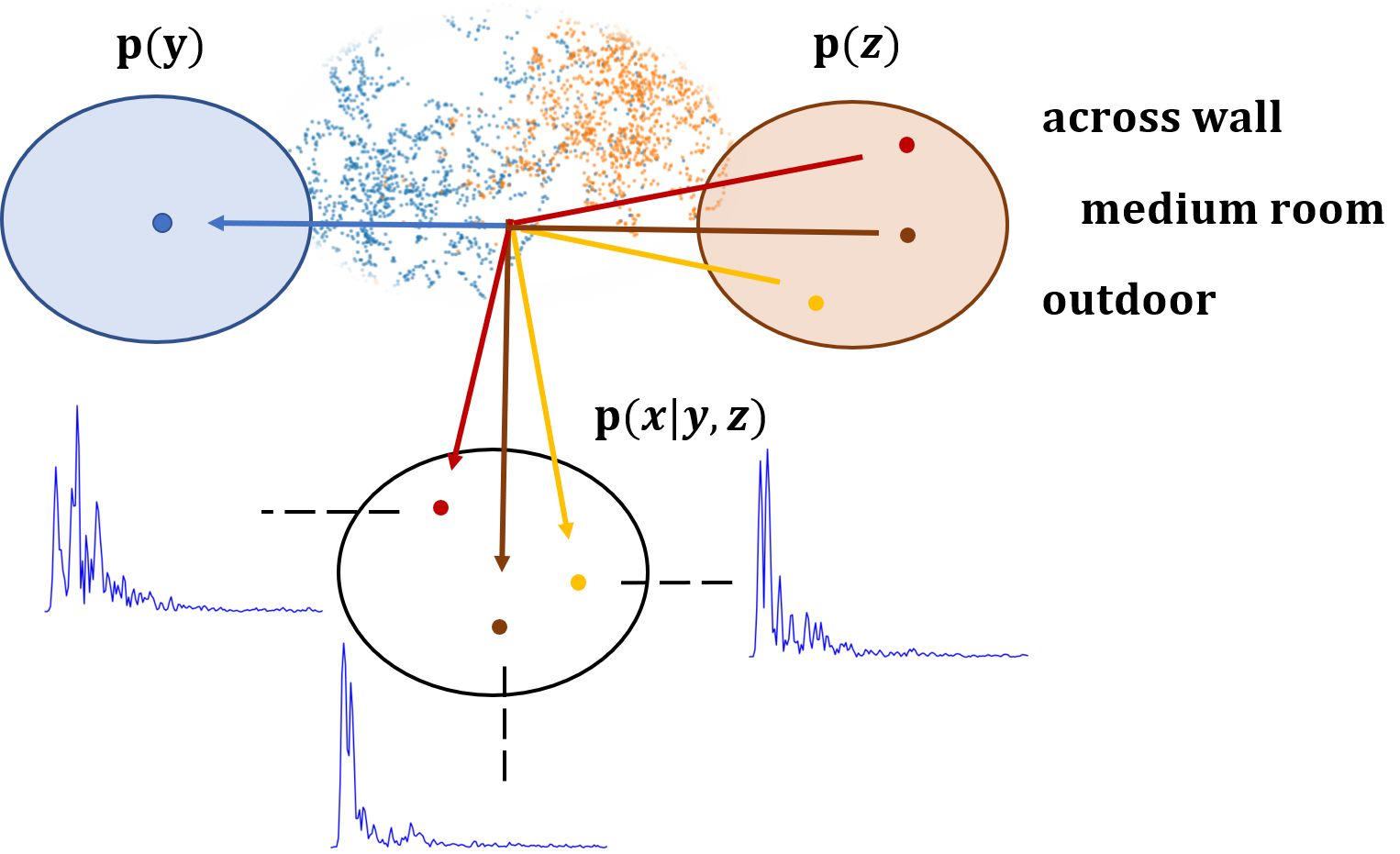}}
        \end{minipage}
        }
        \caption{Illustrations of the proposed Bayesian Model. (a) The graphical model on measured data inference. Dashed lines denote the generative process parameterized by $\boldsymbol{\theta}$, while solid lines denote the embedding from data space to latent space parameterized by $\boldsymbol{\phi}$. (b) waveform samples in data space are generated from according range-related and environment-related features in latent space.}
        \label{fig:graphic}
    \end{center}
\end{figure}

The generative process for received signal is described by a directed probabilistic model in the presence of two unobserved random variables, as illustrated in Fig.\ref{fig:graphic}(a). 
Specifically, signal data $\mathbf{x}$ is assumed to involve two independent latent variables: $\mathbf{y}$ for range-related feature, and $\mathbf{z}$ for range-unrelated features caused by environment variance. The generative process of a signal sample $\mathbf{x}^{(i)}$ can be obtained by first sampling $\textnormal{y}^{(i)}$, $\textnormal{z}^{(i)}$ from $p(\mathbf{y}, \mathbf{z})$, and then generating via the conditional distribution $p(\mathbf{x}|\mathbf{y}^{(i)}, \mathbf{z}^{(i)})$. 
As illustrated in Fig.\ref{fig:graphic}(b), different values environmental variables lead to different output signals.

With the disentanglement of semantics features in the latent space, many measurement-related problems can be solved by mixing the latent variable to incorporate desired semantics, and generate desired results.
In particular, the estimation of ranging error $\Delta d$ can thus be modeled as the estimation of the conditional distribution $p(\Delta d| \mathbf{y})$. 
For such downstream tasks, the disentanglement is required to be achieved. 

\subsection{Variational Lower bound for Data Likelihood}


Suppose data $\mathbf{x}$ are generated by a random process, involving two independent latent variables $\mathbf{y}$ and $\mathbf{z}$. The evidence lower bound (ELBO) $\mathcal{L}$ of the marginal likelihood of data $\mathbf{x}$ can be written as,
\begin{equation}  \label{eq:bound}
    \begin{aligned}
       \log p(\mathbf{x})\geq&\mathcal{L}(q; \mathbf{x})  \\
       =& \mathbb{E}_{q(\mathbf{y}, \mathbf{z}|\mathbf{x})}\big[\log p(\mathbf{x}|\mathbf{y}, \mathbf{z})\big]  \\
       &- \operatorname{D}_{\text{KL}}\big(q(\mathbf{y}|\mathbf{x})\big|\big|p(\mathbf{y})\big) - \operatorname{D}_{\text{KL}}\big(q(\mathbf{z}|\mathbf{x})\big|\big|p(\mathbf{z})\big)
    \end{aligned}
\end{equation}
\noindent where where $\operatorname{D}_{KL}$ is the Kullback-Leibler (KL) divergence, $q(\mathbf{y}|\mathbf{x})$ and $q(\mathbf{z}|\mathbf{x})$ are variational distributions introduced to approximate the true posterior distributions. The inequality is obtained from the Jensen's inequality, achieving equality iff $q(\mathbf{y}|\mathbf{x})=p(\mathbf{y}|\mathbf{x})$ and $q(\mathbf{z}|\mathbf{x})=p(\mathbf{z}|\mathbf{x})$.

Such bound can then be used to find a suitable approximated distribution $q^*$ from some distribution family $Q$ that matches the true distribution $p$ in the generative model, i.e.,
\begin{equation}  \label{eq:opt}
    q^* = \arg \max_{q\in Q} \mathcal{L}(q;\mathbf{x})
\end{equation}

We construct deep neural networks to accumulate knowledge from data and learn these variational distributions.

\section{Semi-Supervised Learning Scheme}
\label{sec:algorithm}

\begin{figure*}[htbp]
      \centerline{
      \includegraphics[width=1.0\textwidth]{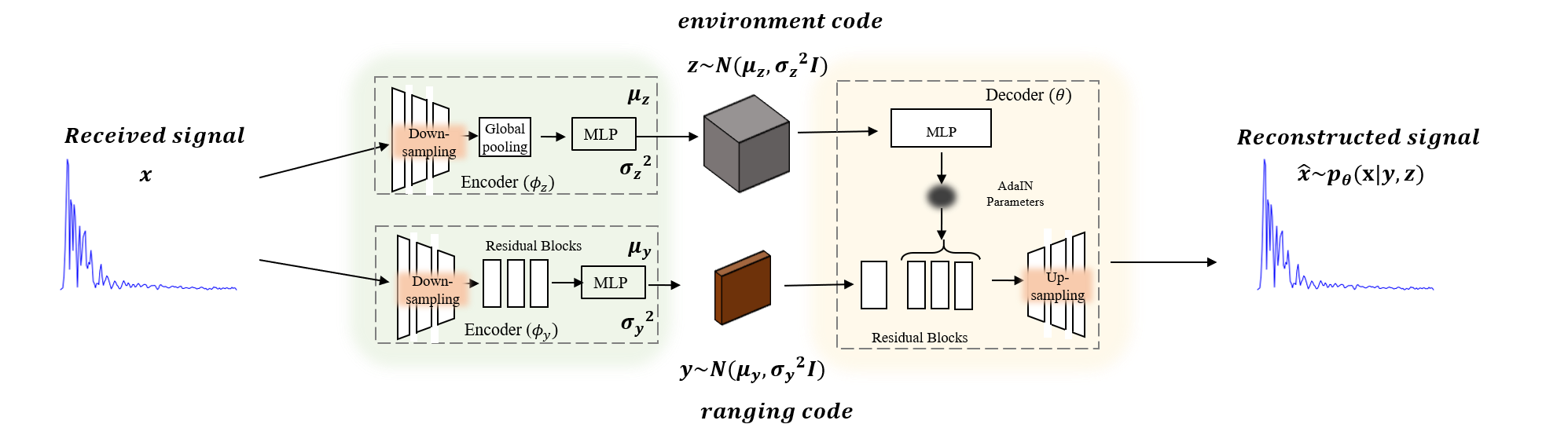}
      }
      \caption{Our auto-encoder architecture for latent variable disentanglement. The encoder for range-related feature consists of several strided convolutional layers followed by residual blocks. The encoder for range-unrelated features constains several strided layers followed by a global average pooling layer and a fully connected layer. The decoder uses a MLP to produce a set of AdaIN \cite{HuaBel:C17} parameters from environment-related features.} 
      \label{fig:arc}
  \end{figure*}
  
In this section, we propose a deep neural network for semi-supervised ranging error mitigation, which enables efficient learning to solve the optimization in equation \eqref{eq:opt}.


\subsection{Parametric Form of the Bound}

The ELBO in \eqref{eq:bound} is transferred into a parametric form to train a specific variational auto-encoder (VAE), which outputs two probabilistic codes for latent variables in the bottleneck. Suppose the likelihood distribution for $\mathbf{x}$ comes from a parametric family of $p_{\boldsymbol{\theta}}(\mathbf{x}|\mathbf{y},\mathbf{z})$ learned by the decoder network $\operatorname{g}_{\text{dec}}(\cdot;{\boldsymbol{\theta}})$ with parameter $\boldsymbol{\theta}$. The variational posterior distribution for latent variables comes from a parametric family of $q_{\boldsymbol{\phi}}(\mathbf{y},\mathbf{z}|\mathbf{x})$ learned by the encoder network $\operatorname{g}_{\text{enc}}(\cdot;\boldsymbol{\phi})$ with parameter $\boldsymbol{\phi}$. Due to the independence between two latent variables, the encoder network can further be decomposed to two sub modules for individual posterior distributions for $\boldsymbol{y}$ and $\boldsymbol{z}$, denoted as $\boldsymbol{\phi}=\{\boldsymbol{\phi}_y, \boldsymbol{\phi}_z\}$.

According to the properties of latent variables, we assume the prior distributions as Gaussian to illustrate randomness:
\begin{equation}  \label{eq:priors}
    \begin{aligned}
        p(\mathbf{y}) &= \mathcal{N}(\mathbf{y}; \boldsymbol{0},\epsilon_y\boldsymbol{I}),  \\
        p(\mathbf{z}) &= \mathcal{N}(\mathbf{z}; \boldsymbol{0},\epsilon_z\boldsymbol{I})
    \end{aligned}
\end{equation}
\noindent where $\epsilon_y$ and $\epsilon_z$ are small values arbitrarily given in practice.

The likelihood distribution for $\mathbf{x}$ is learned by the decoder networks $g(\cdot;{\boldsymbol{\theta}})$ and composes the empirical distribution of $p_{\boldsymbol{\theta}}(\mathbf{x}|\mathbf{y},\mathbf{z})$, i.e.,
\begin{equation} \label{eq:posteriors}
     \mathbf{x} = g(\mathbf{y},\mathbf{z};\boldsymbol{\theta}) \sim p_{\boldsymbol{\theta}}(\mathbf{x}|\mathbf{y},\mathbf{z})
\end{equation}

The approximated posterior distributions are also constructed as Gaussian distributions with mean and variance learned by encoder modules, i.e.,
\begin{equation}
    \begin{aligned}
       q_{\boldsymbol{\phi}_y}(\mathbf{y}|\mathbf{x}) &= \mathcal{N}(\mathbf{y};\hat{\boldsymbol{\mu}}_y, \hat{\boldsymbol{\sigma}}^2_y\boldsymbol{I}) \\
        q_{\boldsymbol{\phi}_z}(\mathbf{z}|\mathbf{x}) &= \mathcal{N}(\mathbf{z};\hat{\boldsymbol{\mu}}_z, \hat{\boldsymbol{\sigma}}^2_z\boldsymbol{I})
    \end{aligned}
\end{equation}
\noindent where $\hat{\boldsymbol{\mu}}_y = \hat{\boldsymbol{\mu}}(\mathbf{x};\boldsymbol{\phi}_y), \hat{\boldsymbol{\sigma}}^2_y=\hat{\boldsymbol{\sigma}}^2(\mathbf{x};\boldsymbol{\phi}_y)$ denote prediction functions for distribution parameters of ranging-related variable $\mathbf{y}$ learned by neural networks with parameter $\boldsymbol{\phi}$, similar for $\hat{\boldsymbol{\mu}}_z$ and $\hat{\boldsymbol{\sigma}}^2_z$ of environment-related variable $\mathbf{z}$. The prior and posterior distributions assumed here are simple and only to illustrate randomness, and can be further assumed if further information for distribution forms of $\mathbf{y}$ and $\mathbf{z}$ are known.

With the aforementioned expression, we derive the parametric version of ELBO w.r.t. signal data $\mathbf{x}$ and parameters $\{\boldsymbol{\phi}_y,\boldsymbol{\phi}_z,\boldsymbol{\theta}\}$ as follows,
\begin{equation}  \label{eq:pa_bound}
    \begin{aligned}
    &\mathcal{L}(\mathbf{x};\boldsymbol{\phi},\boldsymbol{\theta}) = \mathbb{E}_{q_{\boldsymbol{\phi}_y(\mathbf{y}|\mathbf{x})}q_{\boldsymbol{\phi}_z(\mathbf{z}|\mathbf{x})}}\big[\log p_{\boldsymbol{\theta}}(\mathbf{x}|\mathbf{y}, \mathbf{z})\big]  \\
    &- \operatorname{D}_{KL}\big(q_{\boldsymbol{\phi}_y}(\mathbf{y}|\mathbf{x})\big|\big|p(\mathbf{y})\big) - \operatorname{D}_{KL}\big(q_{\boldsymbol{\phi}_z}(\mathbf{z}|\mathbf{x})\big|\big|p(\mathbf{z})\big) 
    \end{aligned}
\end{equation}
\noindent where $\boldsymbol{\phi}=\{\boldsymbol{\phi}_y, \boldsymbol{\phi}_z\}$.

The first term is calculated for efficient error backpropagation as follows,
\begin{equation}  \label{eq:exp}
    \begin{aligned}
        \mathbb{E}_{q_{\boldsymbol{\phi}_y(\mathbf{y}|\mathbf{x})}q_{\boldsymbol{\phi}_z(\mathbf{z}|\mathbf{x})}}\big[\log p_{\boldsymbol{\theta}}(\mathbf{x}|\mathbf{y}, \mathbf{z})\big]
        =\frac{1}{L}\sum_{l=1}^L\log p_{\boldsymbol{\theta}}(\mathbf{x}|\mathbf{y}^{(l)},\mathbf{z}^{(l)})
    \end{aligned}
\end{equation}
\noindent where $\boldsymbol{\epsilon}^{(l)}\sim\mathcal{N}(\boldsymbol{0},\boldsymbol{I})$ is an additional Gaussian distribution, with $l$ the index and $L$ the total number of samples from this distribution. Introduced by the so-called reparameterization trick \cite{KinWel:C13}, $\mathbf{y}^{(l)},\mathbf{z}^{(l)}$ are constructed by the learned mean and variance terms combined with sampling the additional distribution to form the estimation, i.e.,  $\mathbf{y}^{(l)} = \hat{\boldsymbol{\mu}_y} + \hat{\boldsymbol{\sigma}}^2_y \odot \boldsymbol{\epsilon}^{(l)}$, $\mathbf{z}^{(l)} = \hat{\boldsymbol{\mu}_z} + \hat{\boldsymbol{\sigma}}^2_z \odot \boldsymbol{\epsilon}^{(l)}$. 

The last two terms can be calculated analytically with the Gaussian assumptions in equations \eqref{eq:priors}-\eqref{eq:posteriors}. 
Thus the full parametric form for the ELBO in equation \eqref{eq:bound} can be composed, with respect to parameters $\{\boldsymbol{\phi}, \boldsymbol{\theta}\}$ for unknown distributions.

\subsection{Unsupervised Loss Term}
\label{sec:model_unsup}

Given a partially labeled dataset with an unlabeled subset $\mathbb{X}=\{\mathbf{x}^{(i)}\}_{i=1}^M$ of $M$ i.i.d. waveform samples, and a fully labeled subset $(\mathbb{\bar{X}}, \mathbb{\bar{L}})=\{\bar{\mathbf{x}}^{(j)}, \Delta\bar{d}^{(j)}, \bar{k}^{(j)}\}_{j=1}^N$ consisting of $N$ i.i.d. samples with paired signal data $\bar{\mathbf{x}}$, actual ranging error $\Delta \bar{d}$, and the actual label $\bar{k}$ for environment.

The unsupervised loss term is given by the negative form of the parametric ELBO,
utilizing only the waveform data $\mathbf{x}$ without any label information. Such term serves to help disentangle data $\mathbf{x}$ into two independent latent variables. In particular, we construct a modified VAE to formulate the disentanglement in its bottleneck. The unsupervised loss term for such VAE is given as:
\begin{equation}  \label{eq:unsup}
    \mathbb{L}_{unsup}(\mathbb{X},\mathbb{\bar{X}};\boldsymbol{\phi},\boldsymbol{\theta}) = -\sum_{i=1}^M\mathcal{L}(\mathbf{x}^{(i)};\boldsymbol{\phi},\boldsymbol{\theta}) -\sum_{j=1}^N\mathcal{L}(\bar{\mathbf{x}}^{(j)};\boldsymbol{\phi},\boldsymbol{\theta})
\end{equation}


\subsection{Supervised Loss Terms}

The two latent variables could hardly express range-related features and range-unrelated features with only the unsupervised loss term. For further guidance of such disentanglement, the labeled subset $(\mathbb{\bar{X}}, \mathbb{\bar{L}})=\{\bar{\mathbf{x}}^{(j)}, \Delta\bar{d}^{(j)}, \bar{k}^{(j)}\}_{j=1}^N$ is used for an additional supervised loss term. In particular, we construct two additional neural modules to formulate the mappings from latent variables $\mathbf{y}$, $\mathbf{z}$ to $\Delta d$, $k$, respectively. The supervised loss term for sub modules is given as:
\begin{equation}  \label{eq:sup}
    \begin{aligned}
       \mathbb{L}_{\text{sup}}(\mathbb{\bar{X}}, \mathbb{\bar{L}};\boldsymbol{\phi},\boldsymbol{\varphi}) =& \sum_{j=1}^N \Vert \operatorname{f}_{\text{est}}(\operatorname{g}_{\text{enc}}({\mathbf{\bar{x}}}^{(i)};\boldsymbol{\phi});\boldsymbol{\varphi}_e) - \Delta \bar{d}^{(i)}\Vert^2  \\
       & \phantom{\sum_{j=1}^N} + \Vert \operatorname{f}_{\text{cls}}(\operatorname{g}_{\text{enc}}({\mathbf{\bar{x}}}^{(i)};\boldsymbol{\phi});\boldsymbol{\varphi}_c)- {\bar{k}}^{(i)}\Vert^2
    \end{aligned}
\end{equation}
where $\operatorname{f}_{\text{est}}(\cdot;\boldsymbol{\varphi}_e):\mathbf{y}\to \Delta$ maps the range-related features to the real distance label, and $\operatorname{f}_{\text{cls}}(\cdot;\boldsymbol{\varphi}_c):\mathbf{z}\to k$ maps the environment-related features to the environment label. Note that labels $\Delta d$ and $k$ help distinguish the range-related features from range-unrelated feature during the joint training of the neural modules with parameters $\boldsymbol{\theta},\boldsymbol{\phi},\boldsymbol{\varphi}$.

\subsection{Learning Algorithms}
\label{sec:learn}

The VAE guided by the unsupervised term and the two sub modules guided by the supervised term are learned together for fine-tuning parameters, leading to the efficient disentanglement of the range-related features and range-unrelated features. Given a partially labeled dataset $\mathcal{D}=(\mathbb{X},\mathbb{\bar{X}},\mathbb{\bar{L}})$, the total loss function for network learning is thus,
\begin{equation}  \label{eq:func}
    \mathbb{L}(\mathcal{D}; \boldsymbol{\phi},\boldsymbol{\theta},\boldsymbol{\varphi}) = \mathbb{L}_{\text{unsup}}(\mathbb{X},\mathbb{\bar{X}};\boldsymbol{\phi},\boldsymbol{\theta}) + \mathbb{L}_{\text{sup}}(\mathbb{\bar{X}}, \mathbb{\bar{L}};\boldsymbol{\phi},\boldsymbol{\varphi})
\end{equation}
which can lead to different learning results according to the numbers of labeled and unlabeled samples values, according to $M$ and $N$. 



\section{Dataset and Implementations}
\label{sec:dataset}

This section present the UWB dataset for evaluation and the implementation of our algorithm, including network architecture and optimizer for learning.

\subsection{Dataset}

We adopt a public dataset \cite{zenodo} to validate the proposed method. The dataset consists of $100$ data samples in total. Each sample includes a  waveform of $157$ length, an actual ranging error, and two environmental labels for the room setting and blocking materials. The measurements are conducted based on the DWM1000 boards and the actual distances are taken using a Leica AT403 laser tracker to get the ground truth ranging errors. Therefore, the used waveforms and labels for real distance are real world data instead of synthetic data generated by simulation.

The dataset has a great advantage in generality. In particular, measurements are taken in five different room scenarios, including outdoor, big room, medium sized room, small room, and a through-the-wall (TTW) environment. Moreover, obstacles of ten different materials that blocking the LOS path are also taken into account. We assign samples in the medium size room as the testing set ($13210$ samples), and all the other samples as the training set ($36023$ samples), as suggested in \cite{AngMazSalFanChi:J20}.

\subsection{Network Architecture}
\label{sec:data_arc}

Our framework consists of a modified VAE, a classifier, and an estimator, as claimed in Section \ref{sec:learn}. Both the classifier and the estimator are of a simple $3$-layer structure. 
The VAE module inherits a more delicate structure to combine the environment variable and the range variable generically, as illustrated in Fig.\ref{fig:arc}. 

Aside from the structures of the three sub modules, the detailed choices of their layers (i.e., linear, $1$D convolutional, or $2$D convolutional) are discussed in the Section \ref{sec:exp_ab}.

\subsection{Hyper-Parameters}

We use the Adam \cite{DieJim:C14} optimizer with $500$ epochs, where the learning rate is $0.0002$ and decay half every $100$ epochs. The decays of first and second momentum of gradients are $\beta_1=0.5$ and $\beta_2=0.999$, respectively. Weights for the loss terms in \eqref{eq:exp} are set as $\lambda_{\text{unsup}}=10$ and $\lambda_{\text{sup}}=1$ for all our experiments. We build the model in Pytorch \cite{Paszke2017AutomaticDI} and conduct learning on a GTX $1080$ GPU with a memory of $12$ GB and the accelerator powered by the NVIDA Pascal architecture.

\section{Experiments}
\label{sec:exp}


In this section, we evaluate our proposed method on the aforementioned dataset, and validate the robustness of the proposed method under different supervision rate. For convenience of notation, we refer to the proposed semi-supervised method as Semi-VL. 

\subsection{Baselines and Evaluation Metrics}

We adopt three metrics for performance evaluation: the root mean square error (RMSE), the mean absolute error (MAE), and the inference time.

Two learning-based methods are used as the baseline: Support Vector Machine (SVM) method in \cite{MarGifWymWin:J10} and REMNet in \cite{AngMazSalFanChi:J20}. Unlike the proposed method, these methods are trained on a fully labeled dataset. 
Note that results of REMNet are 
directly adopted from its paper, indicating with asterisks next to the values in table~\ref{tab:exp_comp}.

\subsection{Result Analysis}
\label{sec:exp_result}

\begin{table}[t]
\caption{Quantitative results of competitors and the proposed method (referred to as Semi-VL) under different supervision rates.}
\label{tab:exp_comp}
\begin{center}
\begin{small}
\begin{sc}
\begin{tabular}{l|ccr}
\toprule
Methods & RMSE (m) & MAE (m) & Time (ms) \\
\midrule
Unmitigated & 0.1244 & 0.1244 & -  \\
SVM \cite{WymMarGifWin:J12} & 0.1537 & 0.0889 & 0.837  \\
REMnet (LOS) \cite{AngMazSalFanChi:J20} & - & 0.0445* & -  \\
REMnet (NLOS) \cite{AngMazSalFanChi:J20} & - & 0.0687* & -  \\
\midrule
\textit{Semi-VL ($\eta=0.1$)}   & 0.0663 & 0.0176 & 0.242  \\
\textit{Semi-VL ($\eta=0.2$)}   & 0.0603 & 0.0164 & 0.252  \\
\textit{Semi-VL ($\eta=0.4$)}   & 0.0603 & 0.0166 & 0.311  \\
\textit{Semi-VL ($\eta=0.6$)}   & 0.0580 & 0.0163 & 0.210  \\
\textit{Semi-VL ($\eta=0.8$)}   & 0.0567 & \textbf{0.0151} & 0.239  \\
\textit{Semi-VL ($\eta=1.0$)}   & \textbf{0.0558} & 0.0157 & 0.285  \\
\bottomrule
\end{tabular}
\end{sc}
\end{small}
\end{center}
\end{table}

  
Following the semi-supervised dataset claimed in Section \ref{sec:model_unsup}, suppose the dataset consists of $N$ labeled samples and $M$ unlabeled samples, we define the supervision rate $\eta$ as $\eta = \frac{N}{M + N}$. We start by comparing the baseline approaches and the proposed Semi-VL method under $5$ different supervision rates, i.e., $\eta=0.2, 0.4, 0.6, 0.8, 1.0$.

Quantitative results are shown in Table \ref{tab:exp_comp}, including RMSE (m) and MAE (m) for effectiveness and inference time per sample (ms) for efficiency. It can be seen that all the methods successfully mitigate the ranging error to some extend. REMNet outperforms SVM, indicating that learning-based features are superior than hand-crafted features for ranging error mitigation. The proposed Semi-VL achieves significant performance improvement than all the baseline approaches, even at a low rate of supervision. In addition, we compare the results of Semi-VL under different rates of supervision. The results show that the method trained with higher supervision rate tend to have better results, but the performance is rather close. This validate the proposed methodology that the potential of unlabeled data could be excavated by the proposed variational learning method.


\subsection{Convergence of Learning}
\label{sec:exp_curve}

\begin{figure*}[htbp]
    \begin{center}
        \subfigure[]{
        \begin{minipage}[t]{0.45\linewidth}
        \centerline{\includegraphics[width=0.9\textwidth]{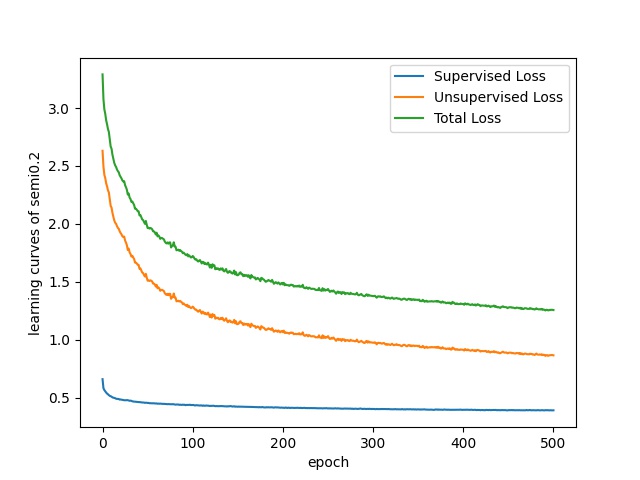}}
        \end{minipage}%
        }
        \subfigure[]{
        \begin{minipage}[t]{0.45\linewidth}
        \centerline{\includegraphics[width=0.9\textwidth]{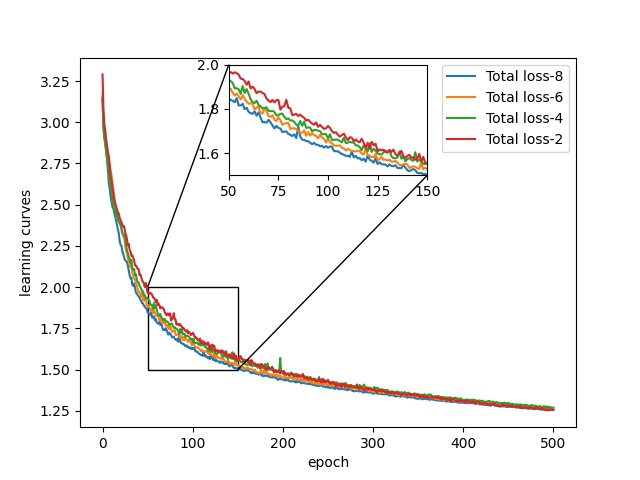}}
        \end{minipage}
        }
        \caption{Learning curves of the proposed Semi-VL. (a) Changes of the total loss, supervised loss, and unsupervised loss terms during the learning process with supervision rate $\eta=0.2$; (b) Learning curves in terms of total loss under different supervision rates.}
        \label{fig:curve}
    \end{center}
\end{figure*}

The learning curves of the proposed Semi-VL under different supervision rates are illustrated in Fig.\ref{fig:curve}. It can be seen that all the learning processes converges at around $300$ epochs, while the stability is positively correlated to the supervision rate.
In particular, the learning process with higher supervision rate is relatively more steady and can converge faster to a desirable result, in accordance with the observations in Section \ref{sec:exp_result}. Moreover, the differences between learning curves with different supervision rates are far from significant. This implies that the proposed Semi-VL can be optimized efficiently and converge easily even at a low supervision rate.

\subsection{Ablation Study}
\label{sec:exp_ab}

\begin{table}[t]
\caption{Quantitative results of the proposed Semi-VL  with different network structures with supervision rate $\eta=1.0$.}
\label{tab:exp_ab}
\begin{center}
\begin{small}
\begin{sc}
\begin{tabular}{l|cccr}
\toprule
\multicolumn{2}{c}{Layer Forms} & RMSE & MAE & Time  \\
\midrule
\multirow{3}{*}{\shortstack{Conv1d \\ AE}}  & Linear Est & 0.0740  & 0.0324 & 0.129  \\
&Conv1d Est   &0.0793 & 0.0394 & 0.156   \\
&Conv2d Est   &0.0786 & 0.0391 & 0.223   \\
\midrule
\multirow{3}{*}{\shortstack{Conv2d \\ AE}}  & Linear Est & \textbf{0.0558}  & \textbf{0.0157} & \textbf{0.285}  \\
&Conv1d Est   &0.0693 & 0.0334 & 0.350   \\
&Conv2d Est   &0.0613 & 0.0274 & 0.378   \\
\bottomrule
\end{tabular}
\end{sc}
\end{small}
\end{center}
\end{table}

Implemented in the architecture described in Section \ref{sec:data_arc}, the choice of detailed layers for the three sub modules remain open. We study three different layers for each module, i.e., linear, $1$D convolutional, and $2$D convolutional layers. The results are obtained under the supervision rate of $1.0$, illustrated in Table \ref{tab:exp_ab}. Note that only one module uses different layer kinds each time, while the other two keep the default setting.

The results show that the combination of a $2$D convolutional VAE, a $2$D convoultional classifier, and a linear estimator achieves the best performance. This may be because the $2$D convolutional operation makes use of the information in correlated sample points in signal and integrates more spatial information. However, such benefit come with a cost of longer inference time. In addition, we observe that the performance of mitigation is not very sensitive to the choice of layers, implying the generality of Semi-VL and the effectiveness of the proposed variational learning methodology.

\section{Conclusion}
\label{sec:con}

We proposed a semi-supervised learning method based on variational Bayes for UWB ranging error mitigation, with two latent variables for range and environment features.
Such methodology can be adapted analogously to multiple signal processing problems. Moreover, the semi-supervised setting introduced here can easily scale to more developed types of supervision.
Our future work will focus on the scalability on supervision types, which is potential to enable transfer learning and a wider scope of applications in wireless communications.

\section*{Acknowledgment}
This research is partially supported by National Key R$\&$D Program of China 2020YFC1511803, the  Basque Government through the ELKARTEK programme, the Spanish Ministry of Science and Innovation through Ramon y Cajal Grant RYC-2016-19383 and Project PID2019-105058GA-I00, and Tsinghua University - OPPO Joint Institute for Mobile Sensing Technology.

\bibliographystyle{IEEEtran}
\bibliography{IEEEabrv,StringDefinitions,SGroupDefinition,refs}

\end{document}